\documentclass[twocolumn,aps,prl,preprintnumbers,showpacs,nofootinbib]{revtex4}
\usepackage[utf8]{inputenc}
\usepackage[english]{babel}
\usepackage{amsmath}
\usepackage{amsfonts}
\usepackage{amssymb}
\usepackage{epsfig}
\usepackage{graphics,psfrag,rotating}
\usepackage{graphicx}
\usepackage{dcolumn}
\usepackage{bm}
\usepackage{epstopdf}
\usepackage{color}
\usepackage[usenames,dvipsnames,svgnames]{xcolor}
\usepackage[colorlinks=true,
            linkcolor=red,
            urlcolor=gray,
            citecolor=blue]{hyperref}
  \usepackage{hyperref}

\newcommand{\lsim}   {\mathrel{\mathop{\kern 0pt \rlap
{\raise.2ex\hbox{$<$}}}
 \lower.9ex\hbox{\kern-.190em $\sim$}}}
\newcommand{\gsim}   {\mathrel{\mathop{\kern 0pt \rlap
{\raise.2ex\hbox{$>$}}}
\lower.9ex\hbox{\kern-.190em $\sim$}}}

\def\3nab{\tilde{\nabla}}

\def\hsp5{\hspace{5mm}}

\def\case#1/#2{\textstyle\frac{#1}{#2}}

\def\ber {\begin{eqnarray}}
\def\eer {\end{eqnarray}}
\def\bea {\begin{eqnarray}}
\def\eea {\end{eqnarray}}

\def\bc {\begin{center}}
\def\ec {\end{center}}
\def\case#1/#2{\frac{#1}{#2}}

\newcommand{\bw}{\begin{widetext}}
\newcommand{\ew}{\end{widetext}}

\newcommand{\be}{\begin{equation}}
\newcommand{\bse}{\begin{subequation}}
\newcommand{\ese}{\end{subequation}}
\newcommand{\ee}{\end{equation}}
\newcommand{\eei}{\end{eqnarray}\indent\indent}
\newcommand{\ba}{\begin{array}}
\newcommand{\ea}{\end{array}}
\newcommand{\bal}{\begin{eqnarray}}
\newcommand{\eal}{\end{eqnarray}}

\def\case#1/#2{\textstyle\frac{#1}{#2} }

\newcommand{\bver}{\begin{verbatim}}
\newcommand{\ever}{\end{verbatim}}

\begin{document}


\title{Does stability in Einstein frame guarantee stability in Jordan frame?}
\author{
Amin Salehi \footnote{Email: salehi.a@lu.ac.ir}
}
\affiliation{Department of Physics, Lorestan University, Khoramabad, Iran}

\date{\today}

\begin{abstract}
 Scalar–tensor theories of gravity can be formulated in the Einstein or in the
Jordan frame, which are related by the conformal transformations. Although the two frames are describe the same physics, and are equivalent, the stability of the field equations in two frames are not the same. Here we implement dynamical system and phase space approach as a robustness tool to investigate this issue.  We concentrate on the Brans-Dicke theory, but the results can easily be generalized. Our analysis show that while there is one-to-one correspondence between critical points in two frames and each critical point in one frame is mapped to its corresponds in other frame , however stability of a critical points in one frame does not grantee the stability in other frame. Hence an unstable point in one frame may be mapped to a stable point in other frame. All trajectories between two critical points in phase space in one frame are different from their corresponds in other ones. This indicates that the dynamical behavior of variables and cosmological parameters are different in two frames.Hence for those features of the study which focus on observational measurements we must use the (JF) where experimental data have their usual interpretation
\end{abstract}


%
%


\maketitle

\section{Introduction}
It is quite evident that the
universe has undergone a smooth transition from a decelerated phase to its present
accelerated phase of expansion \cite{Perlmutter}-\cite{Reiss}. Discovering the source
of cosmic acceleration is one of the biggest challenges of modern cosmology. This remarkable discovery has
led cosmologists to hypothesize the presence of unknown form of energy called dark energy (DE),  which is an exotic matter with negative pressure \cite{copeland}.This surprising
finding has now been confirmed by more recent data coming from
SNeIa surveys
\cite{Knop03,Tonry03,Barris04,Riess04,R06,SNLS,ESSENCE,D07}, large
scale structure \cite{Dode02,Perci02,Szal03,Hawk03,pope04} and
cosmic microwave background (CMBR) anisotropy spectrum
\cite{Boom,Stomp01,Netter02,Rebo04,wmap,WMAP,WMAP3}. All current observations are consistent with a cosmological constant (CC); while this is in some sense the most economical possibility, the CC has its own
theoretical and naturalness problems \cite{Weinberg1}-\cite{Martin}, so it is worthwhile to consider alternatives.
The above observational data properly complete each other and point out
that the dark
energy (DE) is the dominant component of the present universe which occupies about $\%73$ of the energy of
our universe, while dark matter (DM) occupies $\%23$, and the usual
baryonic matter about $\%4$. There are prominent candidates for DE such as the cosmological
constant \cite{Sahni, Weinberg}, a dynamically evolving scalar field ( like quintessence) \cite{Caldwell, Zlatev} or phantom (field with negative energy) \cite{Caldwell2} that explain the cosmic accelerating expansion. Meanwhile, the accelerating
expansion of universe can also be obtained through
modified gravity \cite{Zhu},  brane cosmology and so on \cite{Zhu1}--\cite{set10}. The DE
can track the evolution of the background matter in the
early stage, and only recently, it has negative pressure, and becomes dominant . Thus, its current
condition is nearly independent of the initial conditions \cite{Lyth}--\cite{Easson}.
On the other hand, to explain the early and late time acceleration of the
universe. it is most often the case that such fields interact with matter; directly due to a matter Lagrangian
coupling, indirectly through a coupling to the Ricci scalar or as the result of quantum loop corrections \cite{Damouri}--\cite{Biswass}. If the
scalar field self-interactions are negligible, then the experimental bounds on such a field are very strong; requiring it to either
couple to matter much more weakly than gravity does, or to be very heavy \cite{Uzan}--\cite{Damourm}. Unfortunately, such scalar field is usually very light and its coupling to matter should be tuned to
extremely to small values in order not to be conflict with the Equivalence Principal \cite{nojiri}.
The Brans-Dicke theory of gravity is one of the most popular modified gravity theory which conducted by Brans and Dicke\cite{b1}  and was related with some previous work of Jordan and Fierz \cite{JFBD} for developing an alternative to GR. It is widely used to describe a modification of Einstein's original formulation of General Relativity. This theory can be formulated in the Einstein and
Jordan frame, which are related by the conformal transformations. Although the two frames are describe the same physics, and are equivalent, the stability of the field equations are not the same. Here we implement dynamical system and phase space approach as a robustness tool to investigate this issue.  We concentrate on the Brans-Dicke theory, but the results can easily be generalized.
\section{Mapping between Brans-Dicke, chameleon field and general scalar tensor theory}
 Scalar-tensor theories are usually formulated in two different frameworks, the Jordan Frame (JF) and the Einstein Frame (EF).  It is easier to work in the EF.
We start with the usual Scalar Tensor Theory (STT) action in (JF) \cite{Gilles}
\begin{align}\label{S_JF}
&S={1\over 16\pi G_*} \int d^4x \sqrt{-g}
\Bigl(F(\Phi)~R -
Z(\Phi)~g^{\mu\nu}
\partial_{\mu}\Phi
\partial_{\nu}\Phi\\ \nonumber
&- 2U(\Phi) \Bigr)
+ S_m[\psi_m; g_{\mu\nu}]\ .
\end{align}
Here, $G_*$ denotes the bare gravitational coupling constant , $R$ is
the scalar curvature of $g_{\mu\nu}$, and $g$ its determinant.

The above equations are written in the so-called Jordan frame (JF).
By
conformal transformation of the metric and a redefinition of the
scalar it is possible to obtain field equations in (EF) . Let us call $g^*_{\mu\nu}$ and $\varphi$ the new
variables, and define
\begin{mathletters}
\begin{eqnarray}
g^*_{\mu\nu} &\equiv& F(\Phi)~g_{\mu\nu}\ , \label{g*}\\
\left({d\varphi\over d\Phi}\right)^2 &\equiv& {3\over
4}\left({d\ln F(\Phi)\over d\Phi}\right)^2 + {Z(\Phi)\over
2F(\Phi)}\, \label{varphi}\\
A(\varphi) &\equiv& F^{-1/2}(\Phi)\ ,\label{A}\\
2V(\varphi) &\equiv& U(\Phi)~F^{-2}(\Phi)\ .\label{V}
\end{eqnarray}
\label{2.4}
\end{mathletters}
Action (\ref{S_JF}) then takes the form
\begin{align}\label{S_EF}
&S={1\over 4\pi G_*} \int d^4x \sqrt{-g_*} \left({R^*\over 4} -
{1\over 2} g_*^{\mu\nu} \partial_{\mu}\varphi \partial_{\nu}\varphi
- V(\varphi) \right)\\ \nonumber
&+ S_m[\psi_m; A^2(\varphi)~g^*_{\mu\nu}]\ ,
\end{align}
where $g_*$ is the determinant of $g^*_{\mu\nu}$, $g_*^{\mu\nu}$
its inverse, and $R^*$ its scalar curvature. Note that the above action
looks like the action of chameleon gravity \cite{Hees}  where originally proposed by \cite{Khoury}. Note that matter is
explicitly coupled to the scalar field $\varphi$ through the conformal
factor $A^2(\varphi)$.
 Brans-Dicke theory as a particular case of scalar tensor theory of gravity can be derived by considering,
$F(\Phi) = \Phi$ , $Z(\Phi) = \omega_{BD}/\Phi$ and $ 2ZF+3(dF/d\Phi)^2=2\omega_{BD} +
3$. Thus the field equations in (JF) will be
\begin{eqnarray}\label{fried1}
3H^2=\frac{8\pi G_*\rho}{\Phi}-3H\frac{\dot{\Phi}}{\Phi}
+\frac{\omega_{BD}}{2}\frac{\dot{\Phi}^{2}}{\Phi^{2}}+\frac{U(\Phi)}{\Phi},
\end{eqnarray}
\begin{eqnarray}\label{fried2}
\dot{H}=-\frac{4\pi G_*(\rho+P)}{\Phi}+H\frac{\dot{\Phi}}{2\Phi}
-\frac{\omega_{BD}}{2}\frac{\dot{\Phi}^{2}}{\Phi^{2}}-\frac{\ddot{\Phi}}{2\Phi}
\end{eqnarray}
\begin{eqnarray}\label{phiequatio}
\ddot{\Phi}+3H\dot{\Phi}=\frac{3\Phi(\dot{H}+2H^{2})}{\omega_{BD}}+\frac{\dot{\Phi}}{2\Phi}-\frac{\Phi}{\omega_{BD}}\frac{dU(\Phi)}{ d\Phi}
\end{eqnarray}\begin{eqnarray}\label{phiequation}
\dot{\rho}+3H(\rho+P)=0
\end{eqnarray}

The variables in Brans-Dick theory in (EF) can be related to their corresponding in (JF) as \\
\begin{eqnarray}\label{conformal1}
H_{*} &=&\Phi^{\frac{-1}{2}}(H+\frac{\dot{\Phi}}{2\Phi})\\
\rho_{*}& =&\frac{\rho}{\Phi^{2}}\\
\frac{d\varphi}{dt_{*}}&=&-\frac{\dot{\Phi}}{2\beta\Phi^{\frac{3}{2}}}\label{conformal3}
\end{eqnarray}
Where
\begin{equation}\label{bdp}
\beta=({2\omega_{BD}+3})^{\frac{-1}{2}}
\end{equation}
Hence the field equations in (EF) would be\\
\begin{eqnarray}\label{tm2}
3H^{2}_{*} = 8 \pi G_{*}\rho_*+\dot{\varphi}^{2}+2V(\varphi)
\end{eqnarray}
\begin{eqnarray}\label{tm3}
&&\dot{H_{*}} = -4\pi G_{*}(\rho_*+P_{*})-\dot{\varphi}^{2}\\
&&\ddot{\varphi}+3H_{*}\dot{\varphi}+\frac{d V(\varphi)}{d\varphi} = -4\pi G_{*}\beta(\rho_*-3P_{*})\label{tm4}
\end{eqnarray}

Here, dot denotes derivative respect to $t_{*}$. Note that the field equations (\ref{tm2}) to (\ref{tm4}) are similar to those obtained for chameleon gravity\cite{Hees}.
Here, dot denotes derivative respect to $t_{*}$. Note that the field equations (\ref{tm2}) to (\ref{tm4}) are similar to those obtained for chameleon gravity\cite{Hees}. We also can derive the equations by replacing the physical time $t_{*}$ with the conformal time $\eta_{*}$. Since, $d\eta=\frac{dt}{a}$ and $d\eta_{*}=\frac{dt_{*}}{a_{*}}$, thus,$\mathcal{H}=\frac{dlna}{d\eta}=aH$ and $\mathcal{H_{*}}=\frac{dlna_{*}}{d\eta_{*}}=a_{*}H_*$. Also
\begin{eqnarray}\label{hh}
\mathcal{H}_*=\frac{a'_*}{a_*}=\mathcal{H}-\frac{d \ln (A)}{ d\varphi}\,\varphi^{'}=\mathcal{H}-\beta\varphi^{'},
\label{changedeltaEFJF}
\end{eqnarray}
where prime denotes derivative respect to conformal time $\eta$. The conformal time $\eta$ is the
same in both frames $\eta_{*}\equiv\eta$. Thus, the field equations for Brans-Dicke theory in (EF) will be
\begin{eqnarray}
&3\mathcal{H}_*^2
-\varphi'^2\,=\,2\tilde{\rho_*} +2V(\varphi) a_{*}^2,
\label{g_comps_0}\\
&\mathcal{H}_*^2-\mathcal{H}_*'-\varphi'^2\,=\,  \tilde{\rho_*}(1+c_s^2),
\label{g_comps_j}\\
&\varphi''+2\mathcal{H}_*\varphi'+a_{*}^{2}\frac{dV}{d\varphi} \,=\,- \tilde{\rho_*}(1-3c_s^2)\beta ,
\label{scalar_comps}
\end{eqnarray}
\begin{eqnarray}
\frac{\rho_*'}{\rho_*}=-3\mathcal{H}_*(1+c_s^2)+\beta(1-3c_s^2)\varphi'.
\label{cons_comps}
\end{eqnarray}
Where, $c^{2}_{s}=\frac{P_*}{\rho_*}$ and $\tilde{\rho_*}=4\pi G_{*}\rho_*a_{*}^{2}$. The structure of the field equations is simplified by defining a few variables.
\section{Stability analysis of Brans-Dick theory in (EF)}
In this section we are going to investigate the stability of Brans- Dick theory  in (EF). We consider the power low potential $U(\Phi)=U_{0}\Phi^{m}$ in (JF) which would be mapped to the exponential potentials $V=V_{0}e^{\alpha\varphi}$ in (EF) where $\alpha=2\beta(2-m)$. The
system of equations (\ref{g_comps_0}) to (\ref{cons_comps}) can be transformed to an autonomous system
of differential equations by means of the transformations
 \begin{eqnarray}\label{defe}
\Omega_{1}^{2}=\frac{\tilde{\rho_*}}{3\mathcal{H}_*^{2}},\Omega_{2}^{2}=\frac{\varphi'^{2}}{3\mathcal{H}_*^{2}}
,\Omega_{3}^{2}=\frac{2V(\varphi) a_{*}^{2}}{3\mathcal{H}_*^{2}}
\end{eqnarray}
Equation (\ref{g_comps_0}) gives the following constraint between the variables
\begin{eqnarray}\label{const}
\Omega_{3}^{2}=1-2\Omega_{1}^{2}-\Omega_{2}^{2}
\end{eqnarray}
Now, for the autonomous equations of motions, we obtain
\begin{eqnarray}
\frac{d\Omega_{1}}{dN_{*}}&=&-\frac{1}{2}\left( 1+3c_s^2 \right)\Omega_{1} + \frac{1}{2}\sqrt{3}\beta\left(1-3c_s^2 \right)\Omega_{1}\Omega_{2}\\
&&-\Omega_{1} \left( 1-3\Omega_{2}^2 - 3 \left( 1+3c_s^2 \right)\Omega_{1}^2 \right)\nonumber
\end{eqnarray}
\begin{eqnarray}
\frac{d\Omega_{2}}{dN_{*}}&=&-3\Omega_{2}-\sqrt{3}\beta\left(1-3c_s^2 \right)\Omega_{1}^2
+3\Omega_{2}^{3} +3\Omega_{2}\Omega_{1}^{2}\\
&&+3c_s^2\Omega_{2}\Omega_{1}^{2}-\frac{\alpha}{2}(1-2\Omega_{1}^{2}-\Omega_{2}^{2})\nonumber
\end{eqnarray}
Where $N_{*}=\ln a_{*}$. In order to investigate the evolution of the universe, we need the the essential parameter, $\frac{\mathcal{H}_*^{'}}{\mathcal{H}_*^2}$ . In term of the new variables it would be
\begin{eqnarray}
\frac{\mathcal{H}_*^{'}}{\mathcal{H}_*^2}=1-3(1+c_{s}^{2})\Omega_{1}^{2}-3\Omega_{2}^{2}
\end{eqnarray}
Where, one can obtain the deceleration parameter,$q_{*}$, in (EF) as,
\begin{eqnarray}\label{qe}
q_{*}=-\Big(1+\frac{\dot{H_{*}}}{H_{*}^{2}}\Big)=-\frac{\mathcal{H}_*^{'}}{\mathcal{H}_*^2}=-1+3(1+c_{s}^{2})\Omega_{1}^{2}+3\Omega_{2}^{2}
\end{eqnarray}
\begin{table}
\caption{\label{tmodel} Critical points in (EF) }
\begin{tabular}{cccccc}
Points  &  $\Omega_{1}$  &$\Omega_{2}$ \\
\hline 
\hline
$P_{1}$  &0 & $-\frac{\alpha\sqrt{3}}{6}$ \\
$P_{2}$ & 0 & 1  \\
$P_{3}$ & 0 & -1  \\
$P_{4}$ & $\frac{+ \frac{1}{\sqrt{6}}\left(3c_s^4 - 9\beta^2 c_s^4 -6c_s^2 + 6\beta^2 c_s^2 + 3 - \beta^2 \right)^{\frac{1}{2}}}{ \left( 1-c_s^2 \right)}$ & $-\frac{1}{3}\frac{\left(  3c_s^2 -1 \right)\beta \sqrt{3} }{-1+c_s^2}$  \\
$P_{5}$ & $\frac{- \frac{1}{\sqrt{6}}\left(3c_s^4 - 9\beta^2 c_s^4 -6c_s^2 + 6\beta^2 c_s^2 + 3 - \beta^2 \right)^{\frac{1}{2}}}{ \left( 1-c_s^2 \right)}$ & $-\frac{1}{3}\frac{\left(  3c_s^2 -1 \right)\beta \sqrt{3} }{-1+c_s^2}$  \\
$P_{6}$ & $\frac{1}{2}\frac{(-12+2\alpha^2+6c_s^2\beta\alpha-12c_s^2-2\beta\alpha)^{\frac{1}{2}}}{-\beta+\alpha+3c_s^2\beta} $&$-\frac{\sqrt{3}(1+c_s^2)}{-\beta+\alpha+3c_s^2\beta}$ \\
$P_{7}$ & $\frac{-1}{2}\frac{(-12+2\alpha^2+6c_s^2\beta\alpha-12c_s^2-2\beta\alpha)^{\frac{1}{2}}}{-\beta+\alpha+3c_s^2\beta} $&$-\frac{\sqrt{3}(1+c_s^2)}{-\beta+\alpha+3c_s^2\beta}$ \\
\hline 
\hline\end{tabular}
\end{table}

In the following discussions, we use the Jacobin stability of a dynamical system as the robustness
of the system to small perturbations of the whole trajectory. Jacobin stability analysis offers a powerful and simple method for constraining the
physical properties of different systems, described by second order differential
equations\cite{Sabau}. It is especially important in oscillatory systems where the phase paths
can “spiral in” towards zero, “spiral out” towards infinity, or
reach neutrally stable situations called centers. The eigenvalues of jacobian matrix can be used to determine the stability
of periodic orbits, or limit cycles and predict if the system oscillates near the critical point. In cosmology where
there is the problem of initial conditions, phase space analysis gives us the
possibility of studying all of the evolution paths admissible
for all initial conditions \cite{Salehi1}-\cite{Salehi4}. It is
 useful in visualizing the behavior of the system. In  previous section the critical points of the system have been obtained in term of important parameters$(\beta,\alpha)$. The nature of these points can be determined by the corresponding eigenvalues. Here the eigenvalues of the system are as follows

\begin{align*}
Ev_1=
\begin{bmatrix}
 -3+\frac{\alpha^{2}}{4} \\
 -\frac{3}{2}-\frac{3}{2}c_s^2+\frac{\alpha^{2}}{4}-\frac{1}{4}\beta\alpha+\frac{3}{4}c_s^2\beta\alpha
\end{bmatrix}
\end{align*}


\begin{align*}
Ev_2=
\begin{bmatrix}
 6+\alpha\sqrt{3} \\
 \frac{3}{2}-\frac{3}{2}c_s^2+\frac{1}{2}\beta \sqrt{3} -\frac{3}{2}\beta \sqrt{3}c_s^2
\end{bmatrix}
\end{align*}


\begin{align*}
Ev_3=
\begin{bmatrix}
 6 -\alpha\sqrt{3}\\
 \frac{3}{2}-\frac{3}{2}c_s^2-\frac{1}{2}\beta \sqrt{3}+\frac{3}{2}\beta \sqrt{3}c_s^2
\end{bmatrix}
\end{align*}


\begin{align*}
Ev_4=
\begin{bmatrix}
  -\frac{1}{2} \frac{-3c_s^4 +9\beta^2c_s^4 + 6c_s^2 -6\beta^2c_s^2 -3 +\beta^2}{-1+c_s^2}\\
  -\frac{1}{3}\frac{-18\beta^2 c_s^2-9c_s^4+9+\beta^2+27\beta^2c_s^4+9c_s^2\beta\alpha-3\beta\alpha}{-1+c_s^2}
\end{bmatrix}\\
\end{align*}


\begin{align*}
Ev_5=
\begin{bmatrix}
  -\frac{1}{2} \frac{-3c_s^4 +9\beta^2c_s^4 + 6c_s^2 -6\beta^2c_s^2 -3 +\beta^2}{-1+c_s^2}\\
  -\frac{1}{3}\frac{-18\beta^2 c_s^2-9c_s^4+9+\beta^2+27\beta^2c_s^4+9c_s^2\beta\alpha-3\beta\alpha}{-1+c_s^2}
\end{bmatrix}\\
\end{align*}


\begin{align*}
Ev_6=
\begin{bmatrix}
  \frac{6\beta-18c_s^2\beta+3c_s^2\alpha-3\alpha +\sqrt{D}}{-4\beta+4\alpha+12c_s^2\beta}\\
 \frac{6\beta-18c_s^2\beta+3c_s^2\alpha-3\alpha -\sqrt{A}}{-4\beta+4\alpha+12c_s^2\beta}
\end{bmatrix}\\
Ev_7=
\begin{bmatrix}
  \frac{6\beta-18c_s^2\beta+3c_s^2\alpha-3\alpha +\sqrt{D}}{-4\beta+4\alpha+12c_s^2\beta}\\
 \frac{6\beta-18c_s^2\beta+3c_s^2\alpha-3\alpha -\sqrt{A}}{-4\beta+4\alpha+12c_s^2\beta}
\end{bmatrix}\\
\end{align*}
Where, \\$D=(432-72\alpha^3c_s^2\beta-432c_s^4\beta^2\alpha^2+288c_s^2\beta^2\alpha^2+81c_s^4\alpha^2
+432c_s^2+216c_s^2\beta\alpha-216c_s^2\beta^3\alpha+648c_s^4\beta^3\alpha
-648c_s^6\beta^3\alpha+216c_s^6\beta\alpha-48\beta^2\alpha^2+24\alpha^3\beta+756\beta^2c_s^4-936\beta^2c_s^2-432c_s^4
-432c_s^6+1296c_s^6\beta^2+180\beta^2-108\beta\alpha-63\alpha^2+252c_s^4\beta\alpha+24\beta^3\alpha-18c_s^2\alpha^2).$\\
Generally speaking, the trajectories of the phase space approach to a fixed point if all eigenvalues
get negative values. This fixed point is called stable point, also the trajectories  recede from a fixed point if all eigenvalues have positive values. This fixed point is called unstable point. The fixed points
with both positive and negative eigenvalues are called saddle points, and those trajectories which approach to a saddle
fixed point along some eigenvectors may recede from it along some other eigenvectors.
The behavior of the system near a critical point is spiral if and only if its eigenvalue be complex as $\lambda_{1,2}=\lambda_{r} \pm i\lambda_{I}$.
Because of reality of parameters $\beta$ and $\alpha$, it is obvious that only the eigenvalues $Ev_{6}$ and $Ev_{7}$ can be complex. Thus we can expect the spiral behavior  near the points $p_{6}$ and $p_{7}$.
We investigate the properties of each of the fixed points for the baro tropic equation of state $c_s^2=0$, i.e., dust. \\
\textbf{A:Critical point} $P_{1}$$( \Omega_{1}=0,\Omega_{2}=-\frac{\alpha\sqrt{3}}{6})$. This critical point corresponds to a
solution where the constraint Eqs. (\ref{const}) and (\ref{g_comps_0}) is dominated by \emph{potential-kinetic-scaling solution}. This solution exists for all potentials and only depends on slope of potential $\alpha$. This scaling solution
has two eigenvalues which depend on the slope of potential $\alpha$ and coupling constant $\beta$.
\begin{align}
Ev_1=
\begin{bmatrix}
 -3+\frac{\alpha^{2}}{4} \\
 -\frac{3}{2}+\frac{\alpha^{2}}{4}-\frac{1}{4}\beta\alpha
\end{bmatrix}
\end{align}
The eigenvalue shows that the critical point is stable under the condition \\
$ C I:\left\{
\begin{array}{ll}
\beta<\frac{-6+\alpha^{2}}{\alpha}, -2\sqrt{3}<\alpha<0 \\ \beta>\frac{-6+\alpha^{2}}{\alpha}, 2\sqrt{3}>\alpha>0\\
 \end{array}
\right.
$\\\\
\begin{figure}
\centering
\includegraphics[scale=.3]{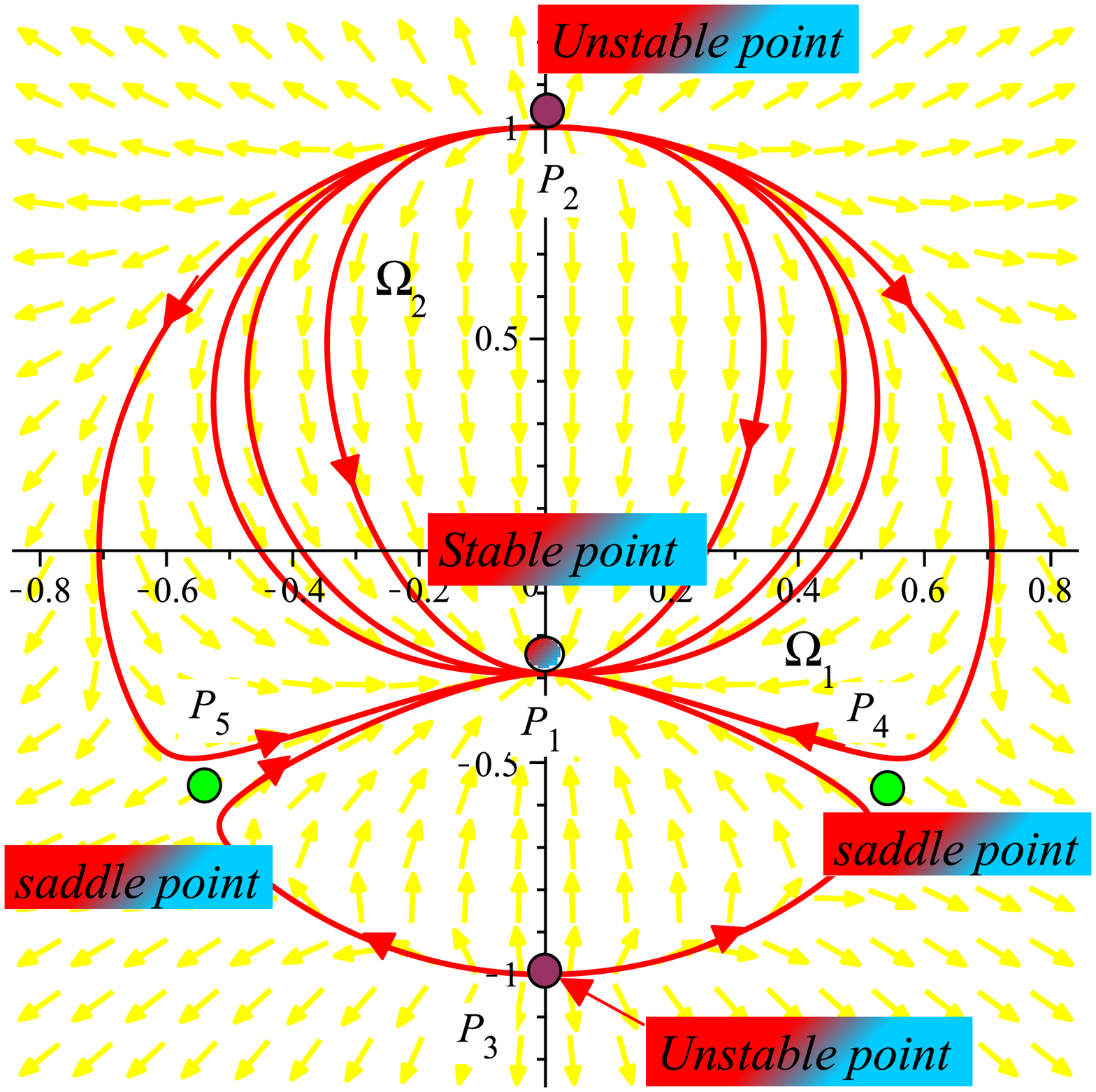}\hspace{0.1 cm}\\
FiG.1. The behavior of the dynamical system in the $\Omega_{1},\Omega_{2}$\\ phase plane for $\beta=1$ and $\alpha=1$. As can bee seen \\ $p_{1}$ is stable, $p_{2}$ and $p_{3}$  are unstable points, $p_{4}$ and $p_{5}$ are saddle points and  $p_{6}$ and $p_{7}$ don't exist
\end{figure}
Fig.1 shows the behavior of the dynamical system in the $\Omega_{1},\Omega_{2}$ phase plane for $\beta=1$ and $\alpha=1$. As can bee seen under the condition $C I $, $p_{1}$ is stable, $p_{2}$ and $p_{3}$  are unstable points, $p_{4}$ and $p_{5}$ are saddle points and  $p_{6}$ and $p_{7}$ don't exist. The non complexity  of the eigenvalues implies that the the system has no spiral behavior near this critical point

\textbf{B:Critical point} $P_{2}$$( \Omega_{1}=0,\Omega_{2}=1)$,corresponds to a \emph{kinetic-scaling solution}. This solution exists for all potentials and is independent of slope of potential $\alpha$ and coupling constant $\beta$. This scaling solution
has two eigenvalues which depend on the slope of potential $\alpha$ and coupling constant $\beta$.
\begin{align}
Ev_2=
\begin{bmatrix}
  6 +\alpha\sqrt{3}\\
 \frac{3}{2}+\frac{1}{2}\beta \sqrt{3}
\end{bmatrix}
\end{align}

\begin{figure}
\centering
\includegraphics[scale=.3]{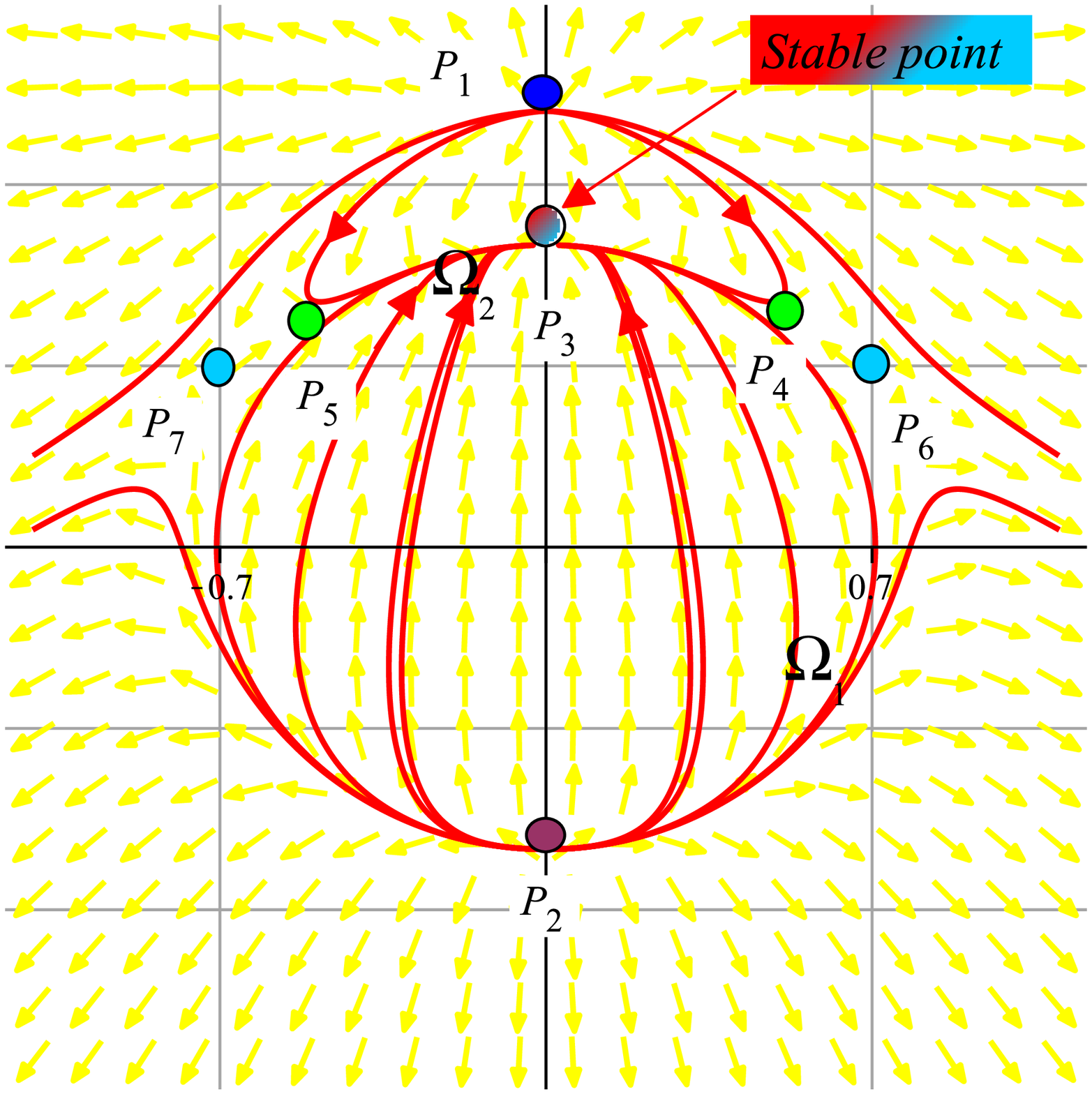}\hspace{0.1 cm}\\
Fig.2.The behavior of the dynamical system in the $\Omega_{1},\Omega_{2}$ phase plane for $\beta=-2$ and $\alpha=-5$. As can be seen  $p_{1}$ and $p_{3}$  are unstable, $p_{2}$ is stable, $p_{4}$ , $p_{5}$,  $p_{6}$ and $p_{7}$ are saddle points.\\
\end{figure}
The eigenvalues show that the critical point is stable for \\
CII:($\beta<-\sqrt{3},\alpha<-2\sqrt{3}$)\\
Fig.2 shows the behavior of the dynamical system in the $\Omega_{1},\Omega_{2}$ phase plane for $\beta=-2$ and $\alpha=-5$. As can be seen  $p_{1}$ and $p_{3}$ are unstable, $p_{2}$ is stable, $p_{4}$ , $p_{5}$,  $p_{6}$ and $p_{7}$ are saddle points.\\
\textbf{C:Critical point} $P_{3}$$( \Omega_{1}=0,\Omega_{2}=-1)$, corresponds to a \emph{kinetic-scaling solution}. This solution exists for all potentials and is independent of slope of potential $\alpha$ and coupling constant $\beta$ however its eigenvalues are depend on slope of potential $\alpha$ and coupling constant $\beta$.
\begin{align}
 Ev_3=
\begin{bmatrix}
  6 -\alpha\sqrt{3}\\
 \frac{3}{2}-\frac{1}{2}\beta \sqrt{3}
\end{bmatrix}
\end{align}

\begin{figure}
\centering
\includegraphics[scale=.3]{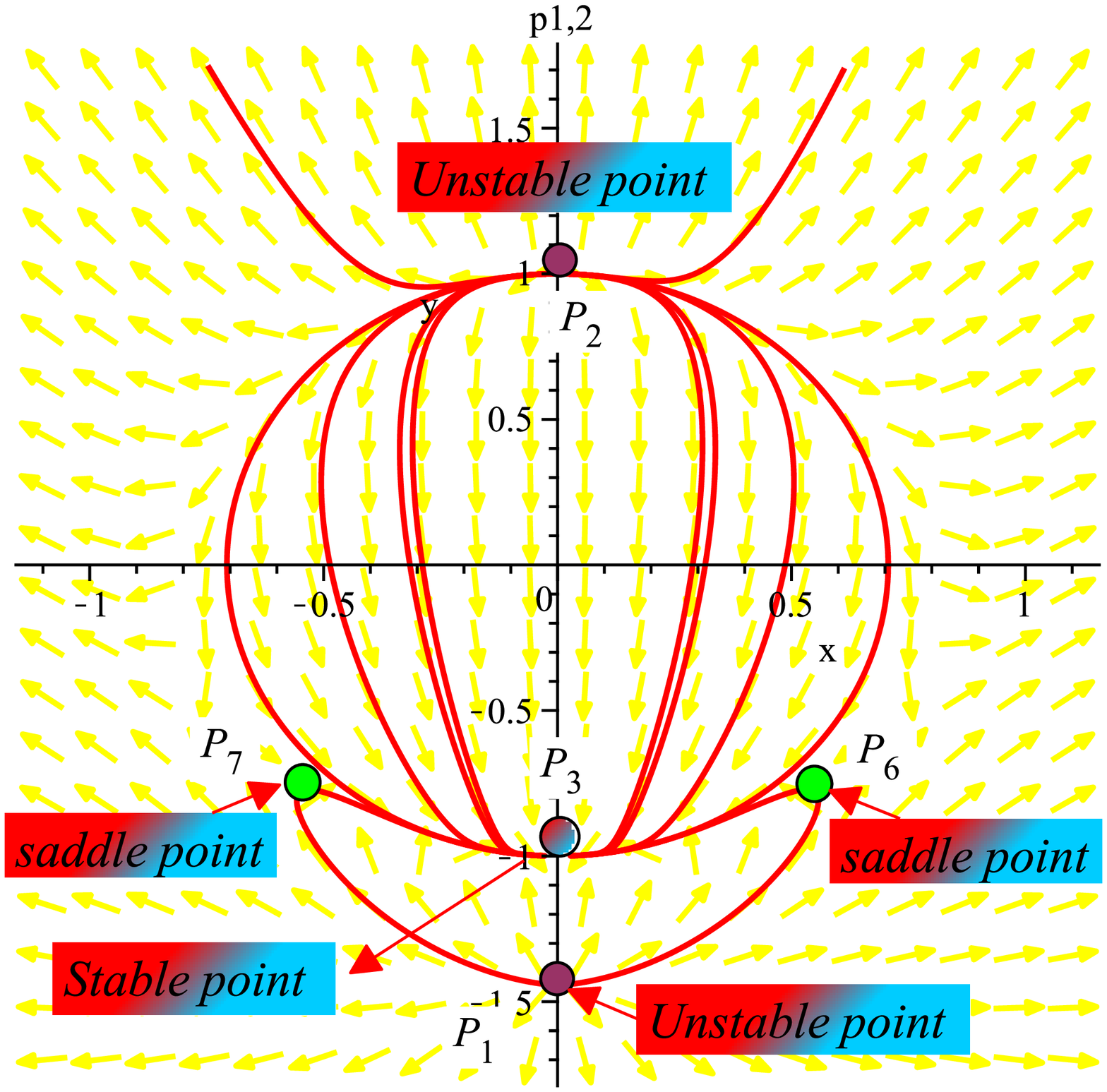}\hspace{0.1 cm}\\
Fig.3.The behavior of the dynamical system in the $\Omega_{1},\Omega_{2}$ phase plane for $\beta=2$ and $\alpha=5$. As can be seen  $p_{1}$ and $p_{2}$  are unstable, $p_{3}$ is stable, $p_{4}$ , $p_{5}$ don't exist  and  $p_{6}$ and $p_{7}$ are saddle points.\\
\end{figure}
The eigenvalues show that the critical point is stable for\\ CIII: ($\beta>\sqrt{3},\alpha>2\sqrt{3}$)\\.Fig.3 shows the behavior of the dynamical system in the $\Omega_{1},\Omega_{2}$ phase plane for $\beta=2$ and $\alpha=5$. As can be seen  $p_{1}$ and $p_{2}$  are unstable, $p_{3}$ is stable, $p_{4}$ , $p_{5}$ don't exist  and  $p_{6}$ and $p_{7}$ are saddle points.\\
\textbf{D:Critical points} $P_{4},P_{5}$$( \Omega_{1}=\pm\sqrt{\frac{3-\beta^{2}}{6}},~\Omega_{2}=-\frac{\sqrt{3}}{3}\beta)$. These critical points are mirror images of each other . These solution exists for $\beta^{2}<3$ and all potentials. The solution
has two eigenvalues which depend on slope of potential $\alpha$ and coupling constant $\beta$.\\
\begin{align}
Ev_{4,5}=
\begin{bmatrix}
 -\frac{3}{2}+\frac{\beta^{2}}{2} \\
 3+\beta^{2}-\beta\alpha
\end{bmatrix}
\end{align}\\
\begin{figure}
\centering
\includegraphics[scale=.3]{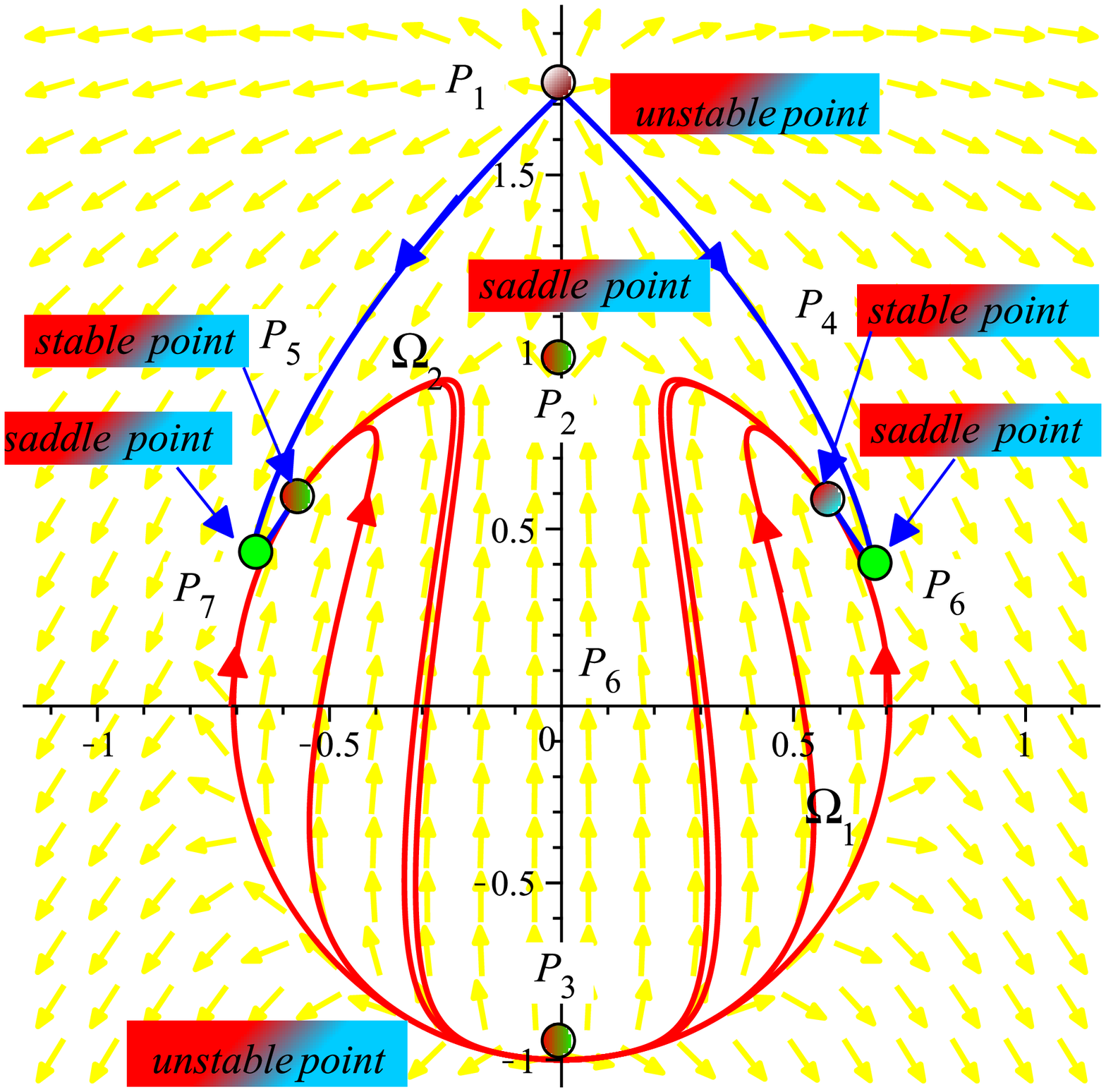}\hspace{0.1 cm}\\
Fig.4.The behavior of the dynamical system in the $\Omega_{1},\Omega_{2}$ phase plane for $\beta=-1$ and $\alpha=-6$. As can be seen  $p_{1}$ and $p_{3}$  are unstable, $p_{2}$, $p_{6}$ and $p_{7}$ are saddle points and  $p_{4}$ and $p_{5}$ are stable points.\\
\end{figure}
The eigenvalues show that the critical point is stable for \\
CIV:$ \left\{
\begin{array}{ll}
\alpha<\frac{3+\beta^{2}}{\beta}, -\sqrt{3}<\beta<0 \\ \alpha>\frac{3+\beta^{2}}{\beta}, \sqrt{3}>\beta>0\\
 \end{array}
\right.
$\\
Fig.4 shows the  behavior of the dynamical system in the $\Omega_{1},\Omega_{2}$ phase plane for $\beta=-1$ and $\alpha=-6$. As can be seen  $p_{1}$ and $p_{3}$  are unstable, $p_{2}$, $p_{6}$ and $p_{7}$ are saddle points and  $p_{4}$ and $p_{5}$ are stable points.\\
\textbf{E:Critical point} $P_{6},P_{7}$($\Omega_{1}= \pm\frac{1}{2}\frac{\sqrt{-12+2\alpha^2-2\beta\alpha}}{-\beta+\alpha}),
\Omega_{2}=\frac{\sqrt{3}}{\beta-\alpha}$).\\ These critical points are mirror images of each other . The solution exists for \\
$ \left\{
\begin{array}{ll}
\beta<\frac{-6+\alpha^{2}}{\alpha}, \alpha>0 \\ \beta>\frac{-6+\alpha^{2}}{\alpha}, \alpha<0\\
 \end{array}
\right.
$\\
The solution
has two eigenvalues which depend on slope of potential $\alpha$ and coupling constant $\beta$.\\
\begin{align}
Ev_{6,7}:
\begin{bmatrix}
  \frac{-6\beta+3\alpha+\sqrt{180\beta^2-108\beta\alpha-63\alpha^2-48\beta^2\alpha^2+24\beta^3\alpha+24\beta\alpha^3+432}}{4(\beta-\alpha)} \\
 \frac{-6\beta+3\alpha-\sqrt{180\beta^2-108\beta\alpha-63\alpha^2-48\beta^2\alpha^2+24\beta^3\alpha+24\beta\alpha^3+432}}{4(\beta-\alpha)}
\end{bmatrix}
\end{align}

\begin{figure}
\centering
\includegraphics[scale=.3]{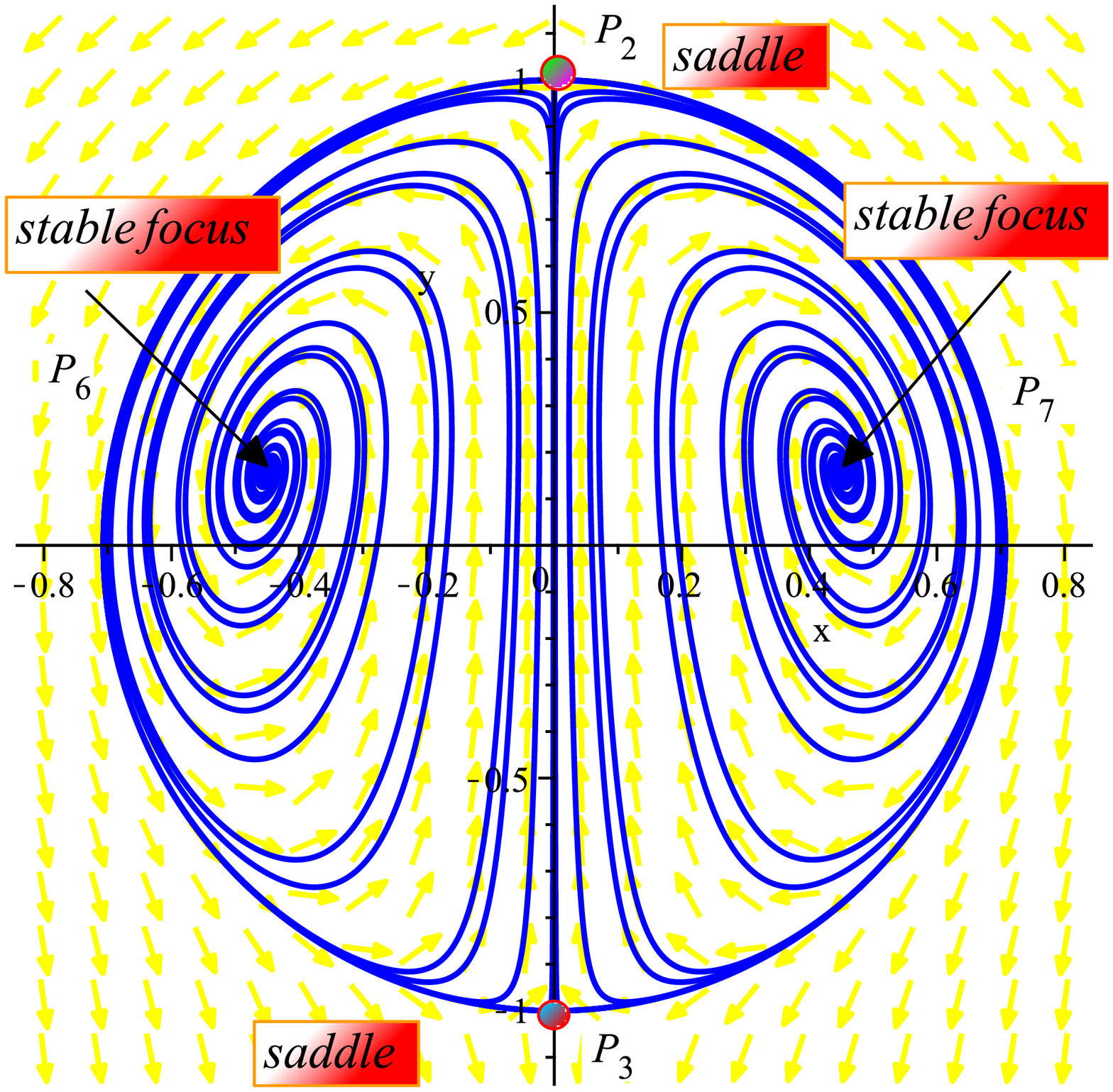}\hspace{0.1 cm}\\
Fig.5.The behavior of the dynamical system in the $\Omega_{1},\Omega_{2}$ phase plane for $\beta=6$ and $\alpha=-5$.\\ As can be seen  $p_{1}$ is unstable $p_{2}$ and $p_{3}$ are saddle points \\ $p_{4}$ and $p_{5}$ dont exist and  $p_{6}$ and $p_{7}$ are stable focus.\\
\end{figure}
Fig.5. shows the behavior of the dynamical system in the $\Omega_{1},\Omega_{2}$ phase plane for $\beta=6$ and $\alpha=-5$. As can be seen  $p_{1}$ is unstable $p_{2}$ and $p_{3}$ are saddle points, , $p_{4}$ and $p_{5}$ don't exist and  $p_{6}$ and $p_{7}$ are stable focus.\\
\begin{figure}
\centering
\includegraphics[scale=.4]{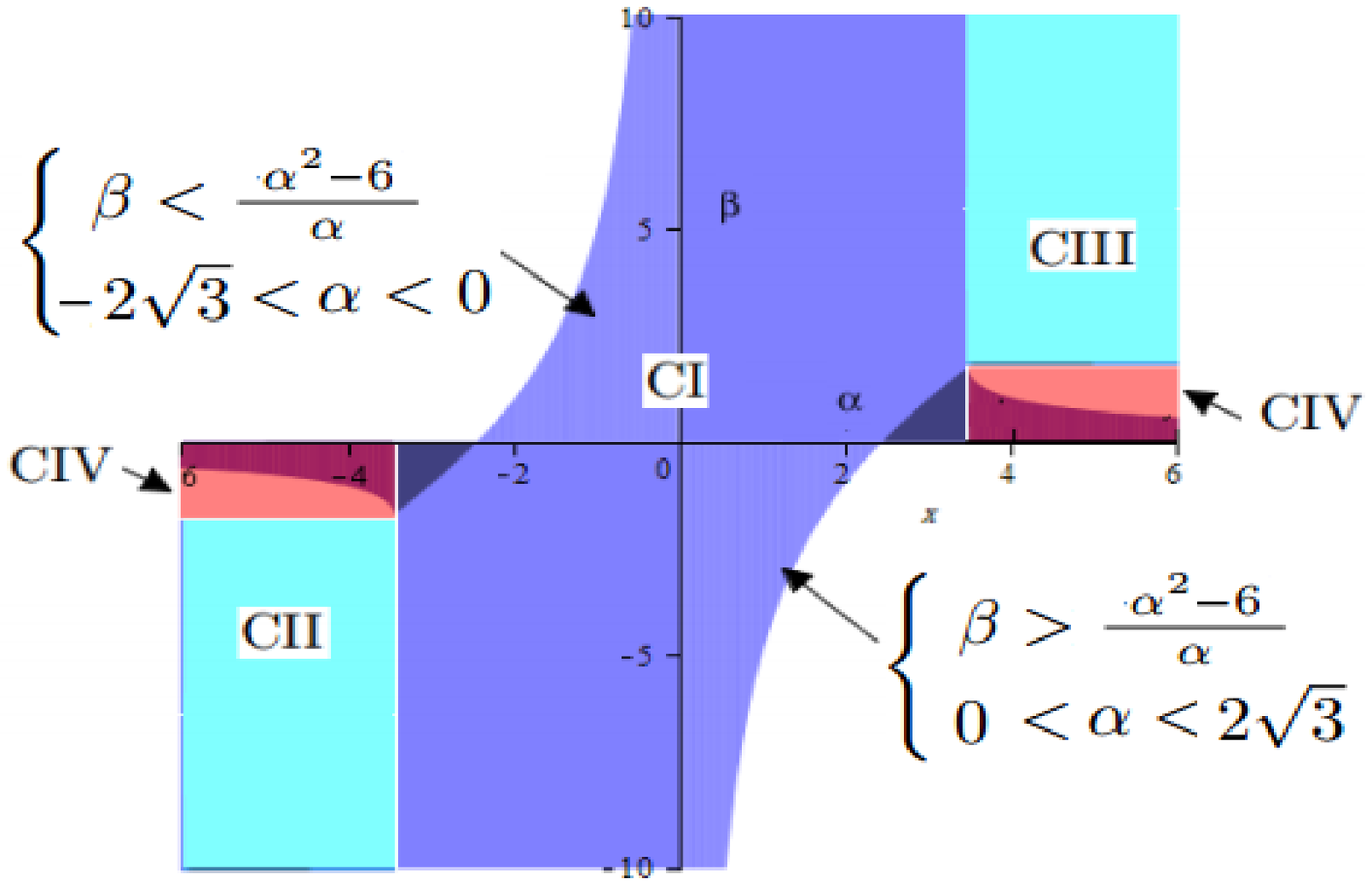}\hspace{0.1 cm}\\
FiG.6. The region of stability for different critical points
\end{figure}
\\

\section{Mapping of stability analysis to (JF) }
In this section using same procedure in (EF), the field equations (\ref{fried1}) to (\ref{phiequation}) can be  transformed to an autonomous system of differential equations by introducing the following dimensionless variables,
\begin{eqnarray}\label{newj}
\Gamma_{1}^{2}=\frac{4 \pi G_{*}\rho}{3H^{2}\Phi},\Gamma_{2}=\frac{\dot{\Phi}}{3\Phi H}
,\Gamma_{3}^{2}=\frac{U(\Phi)}{3\Phi H^{2}}
\end{eqnarray}
However, equations (\ref{conformal1}) to (\ref{conformal3}) are more complicate than equations \ref{g_comps_0}-\ref{cons_comps}, hence in order to derive the autonomous deferential equations in (JF), it is more appropriate to
implement equations (\ref{conformal1}) to (\ref{conformal3}), to make relation between the new variables (\ref{newj}) in (JF)  and variables (\ref{defe}) in (EF) as
\begin{eqnarray}\label{conf1}
&&\Gamma_{2}=\frac{2\beta\Omega_{2}}{\sqrt{3}-3\beta\Omega_{2}}=\frac{2\Omega_{2}}{\sqrt{3}(2\omega_{BD}+3)^{\frac{1}{2}}-3\Omega_{2}}\\
&&\Gamma_{1}=\frac{\sqrt{3}\Omega_{1}}{\sqrt{3}-3\beta\Omega_{2}}=\frac{\sqrt{3}\Omega_{1}(2\omega_{BD}+3)^{\frac{1}{2}}}{\sqrt{3}(2+3)^{\frac{1}{2}}-3\Omega_{2}}\\
&&\Gamma_{3}=\frac{\sqrt{3}\Omega_{3}}{\sqrt{3}-3\beta\Omega_{2}}=\frac{\sqrt{3}\Omega_{3}(2\omega_{BD}+3)^{\frac{1}{2}}}{\sqrt{3}(2\omega_{BD}+3)^{\frac{1}{2}}-3\Omega_{2}}\label{conf3}
\end{eqnarray}
Note that the equations (\ref{conf1}) to (\ref{conf3}) confirm that
\begin{eqnarray}\label{co}
2\Gamma_{1}^{2}-3\Gamma_{2}+\frac{3\omega_{BD}}{2}\Gamma_{2}^{2}+\Gamma_{3}^{2}=1
\end{eqnarray}
Which can be derived from equation (\ref{fried1}) directly. Also
\begin{eqnarray}\label{coo}
\frac{dN_{*}}{dN}=\frac{\mathcal{H}_*}{\mathcal{H}}=\frac{2+3\Gamma_{2}}{2}
\end{eqnarray}
Now, for the autonomous equations of motions in (JF), we obtain
\begin{eqnarray}\label{coo1}
\frac{d\Gamma_{i}}{dN}=\frac{d\Gamma_{i}}{dN_{*}}\frac{dN_{*}}{dN}=\frac{2+3\Gamma_{2}}{2}\frac{d\Gamma_{i}}{dN_{*}}
\end{eqnarray}
Hence using equation (\ref{coo1}) and equations (\ref{conf1}) to (\ref{conf3}), the autonomous equations of motions in (JF) can be related to the corresponding  equations in (EF) as
\begin{eqnarray}\label{au}
&&\frac{d\Gamma_{1}}{dN}=\frac{3(2+3\Gamma_{2})}{2(\sqrt{3}-3\beta\Omega_{2})^{2}}\Big(\frac{d\Omega_{1}}{dN_{*}}-\sqrt{3}\beta(\Omega_{2}\frac{d\Omega_{1}}{dN_{*}}-\Omega_{1}\frac{d\Omega_{2}}{dN_{*}})\Big)\\
&&\frac{d\Gamma_{2}}{dN}=\frac{2+3\Gamma_{2}}{2}\frac{2\sqrt{3}\beta}{(\sqrt{3}-3\beta\Omega_{2})^{2}}\frac{d\Omega_{2}}{dN_{*}}\\
&&\frac{d\Gamma_{3}}{dN}=\frac{3(2+3\Gamma_{2})}{2(\sqrt{3}-3\beta\Omega_{2})^{2}}\Big(\frac{d\Omega_{3}}{dN_{*}}-\sqrt{3}\beta(\Omega_{2}\frac{d\Omega_{3}}{dN_{*}}-\Omega_{3}\frac{d\Omega_{2}}{dN_{*}})\Big)\label{au3}
\end{eqnarray}
Equations (\ref{au}) to (\ref{au3}) indicate that when $(\frac{d\Omega_{1}}{dN_{*}}=\frac{d\Omega_{2}}{dN_{*}}=\frac{d\Omega_{3}}{dN_{*}}=0)$ then their corresponding in (JF) would also be zero $(\frac{d\Gamma_{1}}{dN}=\frac{d\Gamma_{2}}{dN}=\frac{d\Gamma_{3}}{dN}=0)$. This implies that critical points of dynamical system in (EF) would be mapped to their corresponding in (JF) by transformation relations (\ref{conf1}) to (\ref{conf3})(see table.I and II).
\begin{table}
\caption{\label{tmodel} Critical points in (JF) }
\begin{tabular}{cccccc}
Points  &  $\Gamma_{1}$  &$\Gamma_{2}$ \\
\hline 
\hline
$P_{1}$  &0 & $-\frac{2}{3}\frac{\alpha\beta}{\alpha\beta+2}$ \\
$P_{2}$ & 0 & $\frac{2\beta}{\sqrt{3}-3\beta}$  \\
$P_{3}$ & 0 & $\frac{-2\beta}{\sqrt{3}+3\beta}$  \\
$P_{4}$  & $-\frac{\sqrt{18-6\beta^{2}}}{6(\beta^{2}+1}$ & $\frac{-2}{3}\frac{\beta^{2}}{\beta^{2}+1}$ \\
$P_{5}$  & $\frac{\sqrt{18-6\beta^{2}}}{6(\beta^{2}+1}$ & $\frac{-2}{3}\frac{\beta^{2}}{\beta^{2}+1}$ \\
$P_{6}$ & $-\frac{\sqrt{-2\beta\alpha+2\alpha^2-12}}{2(2\beta+\alpha)} $&$-\frac{-2\beta}{(2\beta+\alpha)}$ \\
$P_{7}$ & $\frac{\sqrt{-2\beta\alpha+2\alpha^2-12}}{2(2\beta+\alpha)} $&$-\frac{-2\beta}{(2\beta+\alpha)}$ \\
\hline 
\hline\end{tabular}
\end{table}

 \begin{figure*}
\centering
\includegraphics[scale=.45]{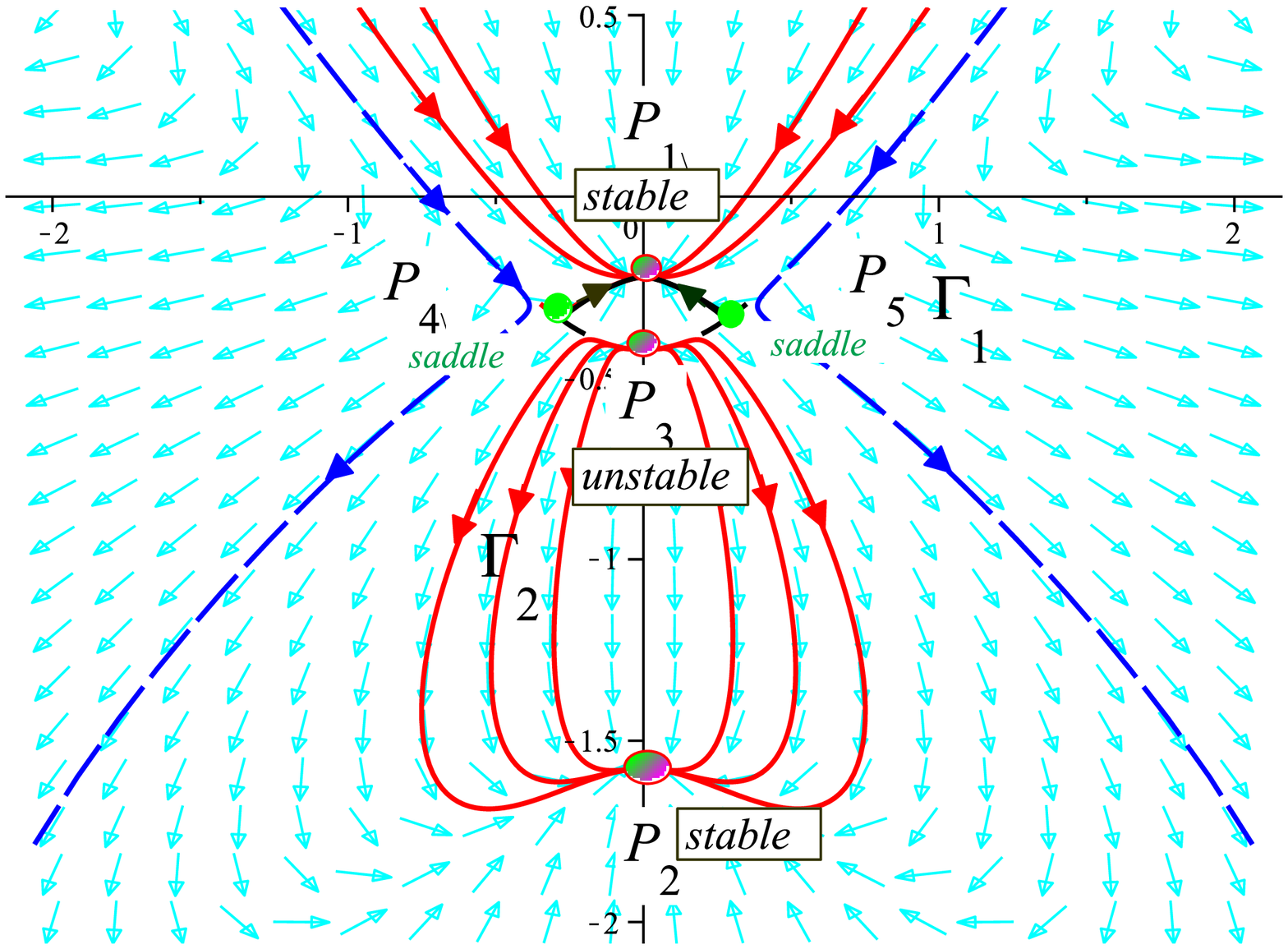}\hspace{0.1 cm}\includegraphics[scale=.33]{p1.eps}\hspace{0.1 cm}\\
\end{figure*}

 \begin{figure*}
\centering
\includegraphics[scale=.8]{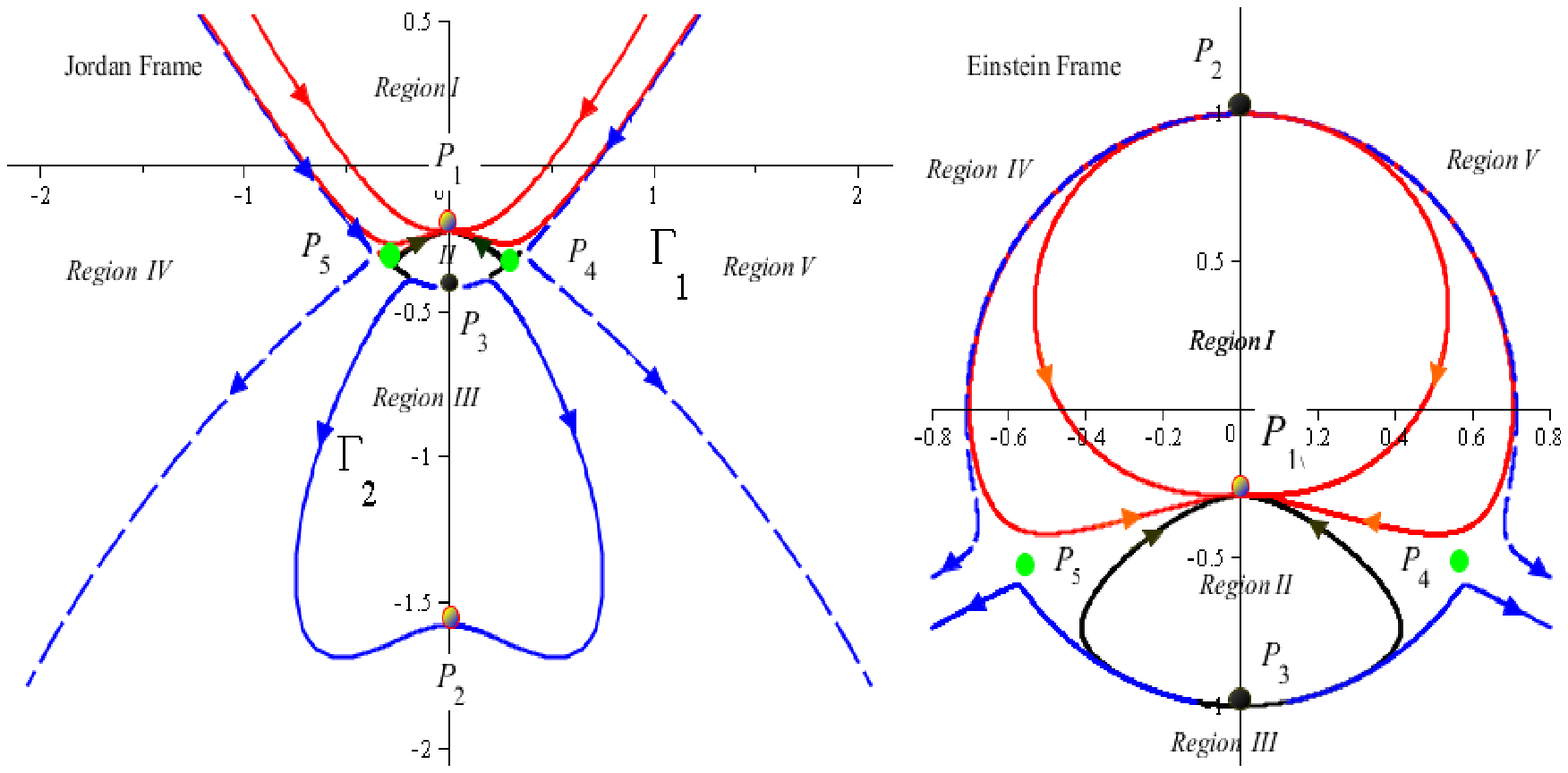}\hspace{0.1 cm}\\
FiG.7. The phase space mapping of (EF) to (JF); The right graph in ($\Gamma_{1},\Gamma_{2}$) phase space in(JF) is mapping of Left ones in ($\Omega_{1},\Omega_{2}$) phase space in (EF) for $\beta=1$ and $\alpha=1$. The lower panel shows the corresponding regions in two frames.
\end{figure*}
Here the eigenvalues of the system are as follows

\begin{align*}
Ev_1=
\begin{bmatrix}
 - \frac{1}{2}\frac{\alpha\beta-\alpha^2+6}{\beta\alpha+2}\\
 \frac{1}{2}\frac{-12+\alpha^2}{\beta\alpha+2}
\end{bmatrix}
\end{align*}
\begin{align*}
Ev_2=
\begin{bmatrix}
 \frac{3}{2}\frac{\sqrt{3}-\beta}{3\beta+\sqrt{3}}\\
 \frac{3(2\sqrt{3}-\beta)}{3\beta+\sqrt{3}}
\end{bmatrix}
\end{align*}
\begin{align*}
Ev_3=
\begin{bmatrix}
 \frac{3}{2}\frac{\sqrt{3}-17\beta-12\alpha\beta^2}{(3\beta+\sqrt{3})}\\
\frac{3(2\sqrt{3}-24\beta+ \alpha-18\alpha\beta^{2})}{(3\beta+\sqrt{3})}
\end{bmatrix}
\end{align*}
\begin{align*}
Ev_4=
\begin{bmatrix}
 \frac{1}{2}\frac{\beta^2-3}{(\beta^{2}+1)}\\
\frac{\beta^2-\alpha\beta+3}{(\beta^{2}+1)}
\end{bmatrix}
\end{align*}
\begin{align*}
Ev_5=
\begin{bmatrix}
 \frac{1}{2}\frac{\beta^2-3}{(\beta^{2}+1)}\\
\frac{\beta^2-\alpha\beta+3}{(\beta^{2}+1)}
\end{bmatrix}
\end{align*}
\begin{align*}
Ev_{6,7}:
\begin{bmatrix}
  \frac{6\beta-3\alpha+\sqrt{180\beta^2-108\beta\alpha-63\alpha^2-48\beta^2\alpha^2+24\beta^3\alpha+24\beta\alpha^3+432}}{4(2\beta+\alpha)} \\
 \frac{6\beta-3\alpha-\sqrt{180\beta^2-108\beta\alpha-63\alpha^2-48\beta^2\alpha^2+24\beta^3\alpha+24\beta\alpha^3+432}}{4(2\beta+\alpha)}
\end{bmatrix}
\end{align*}
As can be seen, while there is one-to-one correspondence between critical points in two frames and each critical point in one frame is mapped to its corresponds in other frame ,  the eigenvalues in (JF) in some critical points are different from those obtained in (EF). This implies that while the critical points in (EF) will be mapped to their corresponding in (JF), however the nature of the critical points may be changed under the transformation and stability of a critical points in one frame does not grantee the stability in other frame.

In Fig.7 the behavior of dynamical system in phase space have been shown in (EF) and its map in (JF)for the same values of $(\alpha=1,\beta=1)$. For this values the critical points in two frames are as;\\
EF$ \left\{
\begin{array}{ll}
P_{1}=(0,-\frac{\sqrt{3}}{6}):stable \\
 P_{2}=(0,1):unstable\\
 P_{3}=(0,-1):unstable\\
 P_{4}=(\frac{\sqrt{3}}{3},-\frac{\sqrt{3}}{3}): saddle \\
 P_{5}=(-\frac{\sqrt{3}}{3},-\frac{\sqrt{3}}{3}):saddle\\
 \end{array}
\right.\\
$JF$ \left\{
\begin{array}{ll}
P_{1}=(0,-0.2):stable \\
 P_{2}=(0,-1.6):stable\\
 P_{3}=(0,-0.4):unstable\\
 P_{4}=(0.3,-0.3):saddle\\
 P_{5}=(-0.3,-0.3):saddle\\
 \end{array}
\right.
$\\
The eigenvalues in (EF) are as follows

\begin{align*}
Ev_1=
\begin{bmatrix}
- \frac{3}{2}\\
\frac{-11}{4}
\end{bmatrix}
,
Ev_2=
\begin{bmatrix}
\frac{3}{2}+\frac{\sqrt{3}}{2}\\
\sqrt{3}+6
\end{bmatrix},
Ev_3=
\begin{bmatrix}
\frac{3}{2}-\frac{\sqrt{3}}{2}\\
-\sqrt{3}+6
\end{bmatrix}
\end{align*}
\begin{align*}
Ev_4=
\begin{bmatrix}
3\\
-1
\end{bmatrix}
,
Ev_5=
\begin{bmatrix}
3\\
-1
\end{bmatrix}
\end{align*}
Their corresponding eigenvalues in (JF) are as follows
\begin{align*}
Ev_1=
\begin{bmatrix}
- 3.2\\
-10.6
\end{bmatrix}
,
Ev_2=
\begin{bmatrix}
-1.\\
-1.8
\end{bmatrix},
Ev_3=
\begin{bmatrix}
2\\
1.6
\end{bmatrix}
\end{align*}
\begin{align*}
Ev_4=
\begin{bmatrix}
1.5\\
-.5
\end{bmatrix}
,
Ev_5=
\begin{bmatrix}
1.5\\
-.5
\end{bmatrix}
\end{align*}
 As can be seen, the critical point $P_{2}$ is unstable in (EF) while its corresponding is stable in (JF). It is also interesting to note that  dynamic of
the deceleration parameters is different in two frames. From equation (\ref{hh}),$\mathcal{H}_*^{'}=\mathcal{H}^{'}-\beta\varphi^{''}$, Hence, the deceleration parameter in (JF) can be derived as
\begin{eqnarray}\label{hh2}
q=-\frac{\mathcal{H}^{'}}{\mathcal{H}^{2}}=-\frac{\mathcal{H}_*^{'}+\beta\varphi^{''}}{(\mathcal{H}_*^{'}+\beta\varphi^{'})^{2}}
=\frac{q_{*}-\beta\frac{\varphi^{''}}{\mathcal{H}_*^{2}}}{(1+\beta\frac{\varphi^{'}}{\mathcal{H}_*})^{2}}
\label{changedeltaEFJF}
\end{eqnarray}
Where using equations (\ref{scalar_comps}) and (\ref{defe}), it will be simplified as,
\begin{eqnarray}\label{hh3}
q=\frac{q_{*}+2\sqrt{3}\beta\Omega_{2}+\frac{\alpha\beta}{2}\Omega_{3}^{2}+3\beta^{2}(1-3c_{s}^{2})\Omega_{1}^{2}}{(1+\sqrt{3}\beta\Omega_{2})^{2}}
\label{changedeltaEFJF}
\end{eqnarray}
 This is an important point to remember: Although
we are looking for cosmological FRW backgrounds whose expansion is accelerating, however, equations (\ref{hh3}) and (\ref{qe}) indicate that, acceleration universe in (JF) may be correspond to deceleration universe in (EF). For example, vanishing potential in (EF) implies that $q_{*}>0$, while  the deceleration parameter $q$ in (JF) may be negative (This can be proved from equations (\ref{qe}), (\ref{const}) and (\ref{hh3})) . As an another straightforward example, at critical point $P_{2}$ in (EF) with ($\Omega_{1}=0,\Omega_{2}=1,\Omega_{3}=0$), the deceleration parameter in (EF) is $q_{*} =2$, while from equation (\ref{hh3}), at this critical point $q=\frac{2+2\sqrt{3}\beta}{(1+\sqrt{3}\beta\Omega_{2})^{2}}$. This indicates that for $\beta<-\frac{\sqrt{3}}{3}$, the deceleration parameter $q<0$.

\section{equivalency of different cosmological models in EF and JF}
Scalar–tensor theories of gravity can be formulated in the Einstein or in the
Jordan frame, which are related by the conformal transformations. Some of the cosmological models can be reconstructed from scalar tensor theories under appropriate conformal metric. As a particular example, we want to discuss equivalency between Brans-Dicke theory and chameleon gravity as the well known models of scalar tensor theories in two different frames. As point out in equation (\ref{S_EF}), it is possible to reconstruct chameleon field equations by transformation of Brans-Dicke equations from Jordan frame (JF) to Einstein frame (EF) under conformal metric $g^*_{\mu\nu}$= $e^{-2\beta\varphi}g_{\mu\nu}$ where $g^*_{\mu\nu}$and $g_{\mu\nu}$ are metrics in Einstein and Jordan Frames respectively and $\beta$ is the chameleon- matter coupling  parameter which would be related to Brans-Dicke parameter $\omega_{BD}$ by $\beta=(2\omega_{BD}+3)^{\frac{-1}{2}}$.
The mathematical equivalency of the models in two different frames has this advantages for our cosmological studies. In principal, for those features of the chameleon study which focus on observational measurements it is more appropriate to use the corresponding Brans-Dicke theory in(JF) where experimental data have their usual interpretation. For example, the consistency between the two theory provides the possibility to derive confidence regions for the value of chameleon-matter coupling constant $\beta$ ( which is still controversial) from corresponding coupling constant $\omega_{BD}$ which severely has been constrained by some observations in (JF) Brans-Dicke theory. Solar System data put very strong constraints on the
$\omega_{BD}$ parameter. The measurement of the Parameterized Post-Newtonian parameter $\gamma$ (see \cite{Will},\cite{Will2}) from the
Cassini mission gives $\omega_{BD}> 40000$ at the $2\sigma$ confidence level \cite{Will2},\cite{Bertotti}. This enable us to find the confidence region for chameleon- matter coupling parameter as $|\beta|<5\times10^{-3}$ in solar system. On cosmological scales, a wide range of values $\omega_{BD}>\{50,2000\}$ have been reported in different studies\cite{Nagata}-\cite{Chen} which determine different confidence region for parameter $\beta$ in cosmological scale. An improvement of pervious studies has been done by\cite{Avilez} using Cosmic
Microwave Background data from Planck. They implemented two types of models. First, the initial
condition of the scalar field is fixed to give the same effective gravitational strength today as
the one measured on the Earth. In this case they find that $\omega_{BD}>692$ at the $(99\% $ confidence level. In the
second type by considering that the initial condition for the scalar is a free parameter they find $\omega_{BD}>890$ at the same confidence
level. These confidence regions for $\omega_{BD}$ put new constraints on parameter $\beta$ as $\beta<0.023$ and $\beta<0.026$ in cosmological scale.\\
However, the important point that we must note is that the evolution of dynamical cosmological parameters such as deceleration parameter which are not equivalent in two frames.

\section{Conclusion}
 In this paper we used dynamical system and phase space approach to show that stability of Brans-Dicke theory in (EF) does not guarantee the stability in (JF).  .We have concentrated on the Brans-Dicke theory, but the results can easily be generalized. Our analysis show that while there is one-to-one correspondence between critical points in two frames and each critical point in one frame is mapped to its corresponds in other frame , however stability of a critical points in one frame does not guarantee the stability in other frame. Hence an unstable point in one frame may be mapped to a stable point in other frame. All trajectories between two critical points in phase space in one frame are different from their corresponds in other ones. This indicates that the dynamical behavior of variables and cosmological parameters are different in two frames. Hence cosmological parameters parameters such as deceleration parameter have different dynamic in two frames where a positive deceleration universe in (EF) may be correspond to an acceleration universe in (JF) and vise versa.

  Hence for those features of the study which focus on observational measurements we must use the (JF) where experimental data have their usual interpretation.
  However we can benefit from equivalency of the equations of two frames. As an particular case we discussed equivalency between Brans-Dicke theory and chameleon gravity as the well known models of scalar tensor theories in two different frames. We explained how we can put constraint on some parameters of chameleon gravity in (EF) using their correspondence in Brans-Dick theory in (JF).

\end{document}